\documentclass[fleqn,10pt]{wlscirep}
\usepackage[utf8]{inputenc}
\usepackage[T1]{fontenc}
\usepackage{xcolor}
\usepackage{tabularray}
\usepackage{soul}

\title{The role of elastic instability on the self-assembly of particle chains in simple shear flow}

\author[1]{Matthew G. Smith}
\author[2]{Graham M. Gibson}
\author[1]{Andreas Link}
\author[3]{Anand Raghavan}
\author[3]{Andrew Clarke}
\author[1]{Thomas Franke}
\author[1,*]{Manlio Tassieri}
\affil[1]{Division of Biomedical Engineering, James Watt School of Engineering, University of Glasgow, Glasgow G12 8LT, U.K.}
\affil[2]{School of Physics and Astronomy, University of Glasgow, Glasgow G12 8QQ, U.K.}
\affil[3]{Schlumberger Cambridge Research, High Cross, Madingley Road, Cambridge, CB3 0EL, U.K.}
\affil[*]{manlio.tassieri@glasgow.ac.uk}

\begin{abstract}
Flow-Induced Self-Assembly (FISA) is the phenomena of particle chaining in viscoelastic fluids while experiencing shear flow. FISA has a large number of applications across many fields including material science, food processing and biomedical engineering. Nonetheless, this phenomena is currently not fully understood and little has been done in literature so far to investigate the possible effects of the shear-induced elastic instability. In this work, a bespoke cone and plate shear cell is used to provide new insights on the FISA dynamics. In particular, we have fine tuned the applied shear rates to investigate the chaining phenomenon of micron-sized spherical particles suspended into a viscoelastic fluid characterised by a distinct onset of elastic instability. 
This has allowed us to reveal three phenomena never reported in literature before, i.e.: (I) the onset of the elastic instability is strongly correlated with an enhancement of FISA; (II) particle chains break apart when a constant shear is applied for `sufficiently' long-time (i.e. much longer than the fluids' longest relaxation time). This latter point correlates well with the outcomes of parallel superposition shear measurements, which (III) reveal a fading of the elastic component of the suspending fluid during continuous shear flows.

\end{abstract}
\begin{document}

\flushbottom
\maketitle
%
%
\thispagestyle{empty}


\section*{Introduction}
Flow-Induced Self-Assembly (FISA) of single particles into long chains while subjected to shear flow is a phenomenon that has been discussed at length since its first description in the $1977$ paper by Michele \textit{et al.}\cite{michele1977}. FISA phenomena occur frequently across a variety of applications, e.g.: (i) in material science, it is well documented that the inclusion of micro and nano particles in polymer melts can greatly enhance the final mechanical properties of products~\cite{vaia2001}; (ii) in food processing, the addition of soft microspheres or microgel droplets can be used to encapsulate phytonutrients for targeted delivery in the gut \cite{shewan2013}; and  (iii) in microfluidics, particle alignment is often required to enhance processes such as counting, analysis and separation \cite{delGiudice2013particle,del2020simultaneous,charjouei2021elasto,hazra2020cross,liu2022self,hu2023self}. 

Currently, the exact mechanism that causes micro-particles to align in simple shear flow is unclear, and the focus of the debate between different schools of thought is mainly on the relative contribution to the driving force governing the phenomena by the elastic and the viscous forces generated during flow, due to the viscoelastic nature of complex fluids. It follows that, most of the arguments have been developed around the relative value assumed by the Weissenberg number ($Wi$), which is a dimensionless parameter used in rheology studies to describe the ratio between the elastic and the viscous forces. For a general overview of the field, an up-to-date review has been masterly drawn by D'Avino and Maffettone \cite{d2015particle}, whose highlights are summarised hereafter for the convenience of the reader. The aforementioned work by Michele \textit{et al.}\cite{michele1977} reported for the first time glass beads forming into long chains and aggregations in a viscoelastic media, under both oscillatory and pipe flows; suggesting that (i) the alignment of particles could be related to the fluid's normal stresses (a measure of the fluid's elastic character) and that (ii) a critical Weissenberg value of $10$ is necessary for the alignment to occur. Subsequent studies by Petit and Noetinger \cite{petit1988shear}, and Lyon \textit{et al}. \cite{lyon2001structure} further corroborated Michele's observation in the case of string formation. Conversely, a more recent study by Won and Kim \cite{won2004alignment} suggests that the shear-thinning nature of the suspension fluid is the driving force for FISA, while normal stresses facilitate migration. Furthermore, Scirocco \textit{et al}. \cite{scirocco2004effect} found that a critical Weissenberg number (as low as $1$) is not solely responsible for string formation as they observed no alignment in Boger fluids (i.e., a viscoelastic fluid with a constant viscosity value). Interestingly, by varying the gap distance between their parallel plate flow cell, Scirocco \textit{et al}. \cite{scirocco2004effect} also found that FISA is a phenomena that occurs within the bulk of the fluid, rather than being a wall effect. However, in contrast to these findings, other studies \cite{pasquino2010effect,pasquino2013migration} observed single particles migration towards the walls, where they would assemble and form strings in the flow direction, when suspended in weakly viscoelastic liquids (i.e., $Wi<<1$). Nonetheless, a recent study by Pasquino \textit{et al}. \cite{pasquino2014} has shown that FISA occurs in both the bulk of the fluid and at the walls of the system; thus implying that such phenomenon is a convoluted function of specific parameters of the system being analysed, such as fluids' viscoelastic properties and flow cell geometries.

Nevertheless, most of the works cited above, and in general in literature, have focused on low Weissenberg numbers (i.e., $Wi<10$) in simple flow cell geometries. Whereas, little has been done to study FISA at relatively high Weissenberg numbers (i.e., $Wi>>10$), where the shear-induced elastic instability may develop. In this regard, it is worth to remind that elastic turbulence, first proposed by Groisman and Steinberg \cite{groisman2000elastic}, has been observed in microchannel flow \cite{zilz2012geometric} and in core-floods \cite{howe2015flow}, and it describes fluctuations in the flow velocity across a broad range of spatial and temporal frequencies, up to a threshold shear rate, where a sharp power-law transition occurs \cite{howe2015flow,pakdel1996elastic,mckinley1996rheological}. This transition is identified by a dimensionless parameter $M_{Norm}$, as posited by Mckinley \textit{et al.}\cite{mckinley1996rheological,pakdel1996elastic}, and further explored in this manuscript for the case of simple shear flow of a shear thinning fluid, which is made of a water based solution of a high molecular weight Polyacrylamide. Interestingly, in a recent study by Howe \textit{et al.}\cite{howe2015flow}, it has been shown that the onset of the shear-induced elastic instability of Polyacrylamide solutions is very distinct and it is concentration independent, but it scales with the square of the polymers' molecular weight.

In this work, a bespoke, counter-rotating cone and plate shear cell has been used to analyse the effects that fluid's viscoelasticity, and more specifically the shear-induced elastic instability, has on FISA. This has been investigated by exploring shear rates that spanned across the onset of elastic instability of a water based solution of Polyacrylamide (PAM), whose frequency-dependent viscoelastic moduli have been determined by means of both classical bulk rheology and microrheology measurements performed with optical tweezers. In agreement with previous works in literature, we have observed that particle chains form in the bulk of the fluid and in the flow direction. In particular, we show that FISA is significantly enhanced by the onset of the elastic instability; although, a significant alignment is also observed at lower shear rates. Moreover, we report evidence of a spontaneous reduction in particles' chain length at relatively long times (i.e., much longer than the fluids' longest relaxation time), which is not associated with particle migration (e.g., sedimentation), but actually to a shear-induced change of the viscoelastic properties of the suspending fluid \cite{tran2023relaxation}. We speculate that these changes may be caused by a shear induced disentanglement of the polymer chains constituting the viscoelastic fluid. This thesis is supported by the outcomes of parallel superposition shear flow measurements, as described below.

\section*{Materials and Methods} 

\subsection*{Particle Suspension}
A dilute solution of particle suspension was prepared by gently mixing $5.2\mu$m diameter polystyrene beads (Bangs Laboratories), at a final concentration of $0.02\%$ w/v, in a water based solution of Polyacrylamide (PAM – molecular weight $18$M, Polysciences Inc.) at a final concentration of $0.07\%$ wt. This concentration is circa ten times higher than the polymer's entanglement concentration, which has been estimated to be (i) $\sim 0.008\%$wt as obtained via viscometry measurements performed by Howe \textit{et al.} \cite{howe2015flow} and (ii) $\sim 0.007\%$wt as read from Figure $6$-(D) of the work by Tassieri \textit{et al.} \cite{Tassieri2019} reporting a comparison between the viscosity values measured by means of multiple rheological techniques. 
Additionally, a $97\%$wt glycerol/water mixture with beads concentration of $0.02\%$ w/v was used as a control. This ratio of glycerol/water mixture was chosen based on the fact that its Newtonian viscosity of $\eta=0.765$ Pa$\cdot$s closely approximates the one of the PAM solution at low frequencies.

\subsection*{Bulk rheology}
Bulk rheology measurements were performed with a stress-controlled rheometer (Netzsch Kinexus Ultra+), utilising a roughened cone and plate geometry ($4^o$ cone, $25$mm radius plate). All measurements were performed at $20^o$C, and prior to a rheological measurement, the sample underwent a pre-shear at $10$s$^{-1}$ for $3$min, to homogenise the sample and remove any history dependant effects from the loading procedure. Partially hydrolysed polyacrylamide (Flopaam $3630$S – molecular weight $18$-$20$MDa, SNF) at a final concentration of $0.07$\%wt in deionised water, without polystyrene beads, was used for all measurements. Parallel superposition measurements were conducted at constant shear rates of $0.49$s$^{-1}$ and $0.79$s$^{-1}$. As the instrument is stress controlled, these shear rates were converted to stress values utilising Carreau – Yasuda fitting parameters from flow curve data. An oscillation stress amplitude of $0.1133$Pa was selected from a dynamic amplitude sweep measurement, lying within the linear viscoelastic region. The parallel superposition measurement was conducted at a frequency of $0.4$Hz, sampling over a duration of $2$hr.

\subsection*{Optical tweezers rig}
Microrheology measurements were performed by using an OT system based on a continuous wave, diode pumped solid state (DPSS) laser (Ventus, Laser Quantum), which provided up to $3$ W at $1,064$ nm. A nematic liquid crystal spatial light modulator (SLM) (BNS, XY series $512\times512$) was used to create and arrange the desired optical trap. The laser entered a custom-made inverted microscope that uses a microscope objective lens (Nikon, $100$x, $1.3$ NA) to both focus the trapping beam and to image the thermal fluctuations of a $4.74\mu$m diameter silica bead (Bangs Laboratories), at room temperature $\sim20^o$C. The sample was mounted on a motorized microscope stage (ASI, MS-$2000$). A complementary metal-oxide semiconductor (CMOS) camera (Dalsa, Genie HM $1024$ GigE) acquired high-speed images of a reduced field-of-view. These images were processed in real-time at up to $\sim 1$ kHz to calculate the center of mass of the bead by using a particle tracking software developed in LabVIEW (National Instruments), running on a standard desktop PC \cite{bowman2014red,gibson2008measuring}.

\subsection*{A Theoretical Background of Microrheology with Optical Tweezers}

Microrheology is an branch of Rheology (the study of the flow of matter), and it is focused on the characterisation of the viscoelastic properties of complex fluids by using sample volumes in the micro-litre range; thus making microrheological methods ideal candidates for measuring rare or precious samples, with a clear advantage over classical bulk rheology approaches that require millilitres of sample volume. Microrheology techniques are categorised into either ``passive'' or ``active'' depending on whether the tracer particle, suspended in the target fluid, is driven by thermal fluctuations within the fluid, or by means of an external force, respectively.

Developed in the 1970s \cite{Ashkin1970}, optical tweezers (OT) utilise a monochromatic laser beam, focused through a microscope objective with a high numerical aperture, to optically trap in three dimensions a micron sized particle, suspended in a fluid; a schematic representation is presented in Fig\ref{fig:MicroPAM}-(A). Once trapped, the particle `feels' a harmonic potential, therefore the restoring force exerted on the particle is linearly proportional to the distance from the center of the trap, provided the displacement is within the bead diameter, and it is of the order of a few $\mu$N. In this work, passive microrheology with OT (MOT) has been used to measure the viscoelastic properties of a PAM solution at a concentration of $0.07\%$ wt. The thermal fluctuations of an optically trapped particle were analysed by means of the theoretical framework developed by Tassieri \textit{et al}.\cite{Tassieri2015_MOT, tassieri2016microrheology,Tassieri2019}, which is here summarised for convenience of the reader.

The Brownian motion of an optically trapped particle can uncover the viscoelastic properties of the suspending fluid when its trajectory is analysed by means of a generalised Langevin equation (Eqn.\ref{eqn:Langevin}) -- as first established by Mason and Weitz \cite{Mason1995} for the case of a freely diffusing particle -- which in this case reads: 
\begin{equation}
    m\vec{a}(t)=\vec{f_R}(t)-\int_{0}^{t}\xi(t-\tau)\vec{v}(\tau)d\tau-\kappa\vec{r}(t),
    \label{eqn:Langevin}
\end{equation}
where $m$ is the mass of the particle, $\vec{a}(t)$ is its acceleration, $\vec{v}(\tau)$ is its velocity, $\vec{r}(t)$ is its position, $\vec{f_R}(t)$ is the Gaussian white noise term used for modelling the stochastic forces acting on the particle, and $\xi(t)$ is the generalised time-dependent memory function accounting for the viscoelastic nature of the fluid.

As described by Tassieri \textit{et al}.\cite{Tassieri2015_MOT, tassieri2016microrheology,Tassieri2019, Smith2021}, Eqn.\ref{eqn:Langevin} can be solved for the fluid's complex modulus ($G^*(\omega)$) \textit{via} either the normalised mean squared displacement (NMSD) $\Pi(\tau)=\langle\Delta r^2(\tau)\rangle / 2\left\langle r^2\right\rangle$ or the normalised position autocorrelation function (NPAF) $A(\tau)=\left\langle \vec{r}(t)\vec{r}(t+\tau)\right\rangle / \left\langle r^2\right\rangle$, which are simply related to each another as:
\begin{equation}
    \Pi(\tau)=\frac{\langle \Delta r^2(\tau)\rangle_{t_0}}{2\langle r^2\rangle_{eq.}}\equiv \frac{\langle [r(t_0+\tau)-r(t_0)]^2\rangle_{t_0}}{2\langle r^2\rangle_{eq.}} =1-A(\tau),
    \label{eqn:NMSD}
\end{equation}
where $\tau$ is the lag-time ($t-t_0$) and the brackets $\langle...\rangle_{t_0}$ represent an average over all initial times $t_0$.
The fluid's complex modulus can then be expressed as:
\begin{equation}
\label{eqn:G*OT}
	G^*(\omega)\frac{6\pi a}{\kappa}=\left(\frac{1}{i\omega \hat{\Pi}(\omega)}-1\right)\equiv \left(\frac{1}{i\omega\hat{A}(\omega)}-1\right)^{-1}\equiv \frac{\hat{A}(\omega)}{\hat{\Pi}(\omega)}
\end{equation}
where $G^*(\omega)$ is a complex number, whose real and imaginary parts define the elastic ($G'(\omega)$) and the viscous ($G''(\omega)$) moduli of the fluid, respectively; $a$ is the particle radius, $\kappa$ is the optical trap stiffness, $\hat{\Pi}(\omega)$ and $\hat{A}(\omega)$ are the Fourier transforms of the NMSD and the NPAF, respectively. To obtain Eqn.\ref{eqn:G*OT}, the inertial term ($m\omega^{2}$) present in the original works\cite{Tassieri2010,Preece2011} has been here neglected, because for micron-sized particles it only becomes significant at frequencies of the order of MHz. From Eqn.\ref{eqn:G*OT} it is a trivial step to derive the complex viscosity ($\eta^*(\omega)$) of the fluid:
\begin{equation}
    |\eta^*(\omega)| = \frac{\sqrt{G'^2(\omega)+G''^2(\omega)}}{\omega}.
\end{equation}

\begin{figure}[t!]
    \centering
    \includegraphics[trim={0 0 0 0},clip,width=\textwidth]{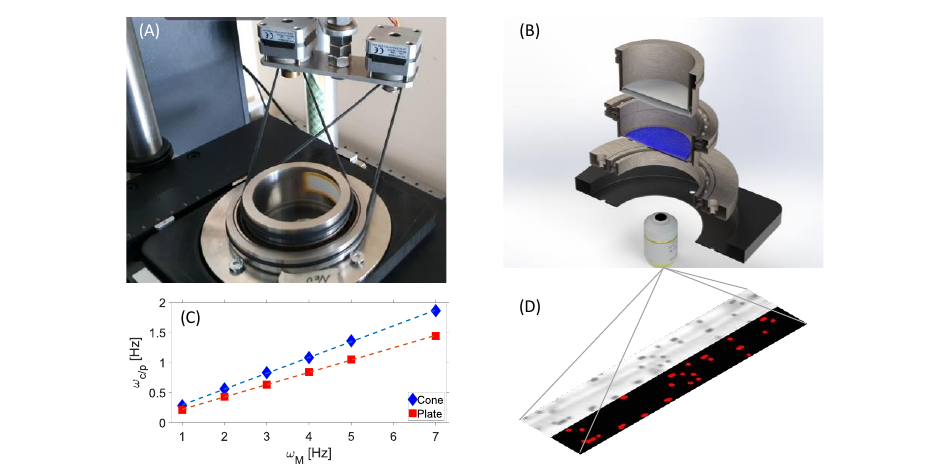}
    \caption{(A) A picture of the bespoke shear cell mounted on a microscope stage with two driving motors. (B) An exploded schematic representation of the shear cell showing the transparent cone (grey area) and plate (blue area) geometries. (C) Calibration curve comparing the rotational frequency of the cone and the plate ($\omega_{c/p}$) versus the rotational frequency of the motors ($\omega_M$). The conversion factor for the cone and the plate was $0.263$ and $0.205$, respectively. (D) Example of a typical frame captured during experiments and the same image post processing. In red are the single beads, dimers and one trimer identified by using the particle tracking software developed in LabVIEW (National Instruments) for this work, running on a standard desktop PC.}
    \label{fig:Calibration}
\end{figure}

\subsection*{Shear Cell}
The shear cell used in this work had a bespoke cone and plate design, both of which were transparent allowing the use of an optical microscope to capture images and the generation of a uniform shear rate along the diameter of the shear cell. An image and an exploded schematic representation of the setup is shown in Fig.\ref{fig:Calibration}-(A,B). The shear cell is positioned on top of a microscope stage and is composed of several individual parts including: the 3D printed base, the shear cell mount, the plate body, and the cone, which slot together creating a chamber for the fluid being analysed. As shown in Fig.\ref{fig:Calibration}-(B), the base of the shear cell (black) is 3D printed that allows the setup to fit within the microscope stage and hence accurately position the shear cell within the optical path of the microscope. The mount for the shear cell is a ring that screws into the 3D printed base and houses a bearing around its internal diameter. On this bearing sits the plate body allowing the plate to rotate freely. The plate body also has a bearing on which the cone body sits. The fluid is applied between the cone and the plate, which in Fig.\ref{fig:Calibration}-(B) is schematically represented by the blue area. The rotation of the cone and the plate was driven by two independent stepper motors via two nitrile O-rings as shown in Fig.\ref{fig:Calibration}-(A). Motor control was provided by two individual Arduino boards, each driving an A4988 stepper motor controller, interfaced with a Labview program that allowed us to control motor speed. 

Prior to performing the measurements, a calibration of the relative speed between the two electric motors and the related cone and plate geometries was performed.
This consisted in varying the rotational frequency of the motors and measuring those of the related geometries, which was much lower because of the relatively high gear ratio.
The rotational frequency of the driven cone/plate ($\omega_{c/p}$) can be calculated using the Eqn.\ref{eqn:RotatFreq}, which is based on the diameter ratio between motor and the cone/plate, $d_m$ and $d_{c/p}$ respectively.

\begin{equation}
    \omega_{c/p} = \frac{d_m}{d_{c/p}}\omega_M,
    \label{eqn:RotatFreq}
\end{equation}
where $\omega_M$ is the rotational frequency of the motor shaft and the diameter of the cone and the plate are $60$ and $80$mm respectively. 

Calibration was measured in three configurations: (i) with either the cone or the plate stationary, (ii) with both rotating in the same direction and (iii) with both rotating in opposite directions. The purpose of measuring in each configuration was to make sure that the rotation of the cone did not influence the rotation of the plate and \textit{vice versa}. In each configuration the time required for the cone and the plate to complete 10 revolutions was measured. The rotational frequency of both the cone and the plate were graphed against the rotational frequency of the motors, shown in Fig.\ref{fig:Calibration}-(C), and the gradient of the line is the inverse of the gear ratio.

 The shear cell was counter rotated, and the shear rates explored during the measurements are those reported in Table\ref{tab:ShearRate}, which shows also the rotational frequency of the cone driving motor ($\omega_{cM}$) and the plate driving motor ($\omega_{pM}$) required to achieve an equal rotational frequency of the cone and the plate ($\omega_{c/p}$) for each of the shear rate examined:

\begin{equation}
    \dot{\gamma} = \frac{(\omega_{c}+\omega_{p})}{tan(\alpha)} \simeq \frac{(\omega_{c}+\omega_{p})}{\alpha},
\end{equation}

where $\dot{\gamma}$ is the shear rate and $\alpha$ is the angle of the cone (i.e.$4.20^\circ\pm 0.25^\circ$)~\cite{derks2004confocal,tees1993interaction}. The truncation gap of the cone was $125\mu m$, which is significantly larger than the polystyrene beads used in the suspension (diameter of $5.2\mu m$) and therefore gap effects are negligible~\cite{scirocco2004effect}.

\begin{table}[ht]
\centering
\begin{tabular}{|c|c|c|c|c|}

\hline
$\dot{\gamma }$ $[s^{-1}]$ & $\omega_{c}$ [$rad\cdot s^{-1}$] & $\omega_{p}$ [$rad\cdot s^{-1}$] & $\omega_{cM}$ [$rad\cdot s^{-1}$] & $\omega_{pM}$ [$rad\cdot s^{-1}$] \\
\hline
27.682 & 1.025 & 1.005 & 3.896 & 4.901 \\
\hline
43.479 & 1.603 & 1.584 & 6.095 & 7.728 \\
\hline
68.403 & 2.528 & 2.486 & 9.613 & 12.127 \\
\hline
107.745 & 3.983 & 3.916 & 15.143 & 19.101 \\
\hline
169.028 & 6.246 & 6.144 & 23.750 & 29.971 \\
\hline
265.690 & 9.816 & 9.660 & 37.322 & 47.124 \\
\hline

\end{tabular}
\caption{\label{tab:ShearRate} Table of shear rates and rotational frequencies explored in this work.}
\end{table}

The sample, $750\mu$l of $0.07\%$wt PAM-bead solution, was inserted between the cone and the glass coverslip (plate) and then a constant shear rate was applied for $30$ minutes.
Images were acquired using a Teledyne Dalsa Genie camera at $960$fps with an exposure time of $400\mu$s. The high frame rate was achieved by reducing the region of interest of the camera from $1400 \times 1024$ pixels to $752 \times 100$ pixels. The window was aligned in the direction of flow, so that chaining could be observed through the major axis of the window and the focal plane was positioned in the center of the shear flow, which was kept constant for each experiment. A Labview program has been developed to allow us to run the camera at $960$fps, but capture images at set intervals in milliseconds, thus improving significantly the performance of the image acquisition and reducing the time taken for analysis. The acquisition rate of the Labview program was set to $1.2$s for all experiments, thus returning $1500$ frames for each measurement; which were performed in triplicates. Fig.\ref{fig:Calibration}-(D) shows a typical image frame captured both before and after image processing.

Image analysis was carried out by using custom code, developed using National Instruments Labview, to automatically process the acquired images. Each frame was analysed by first removing the background of the image, as shown in Fig.1-(D), and then each particle/chain was identified automatically using functions from the National Instruments, Vision Development Module, which can discriminate different objects via their (i) area (as each additional bead in a given chain increases the object area linearly) and (ii) elongation value (which regulated the counting of only chains rather than artifacts such as particles' agglomerates); thus, allowing us to classify whether an identified object belonged to a specific chain length (i.e., single bead, dimer, trimer, tetra or penta). Once classified, the total number of each chain length per frame was exported into a spreadsheet.

\section*{Results and Discussion}

\subsection*{Rheological characterization of a Polyacrylamide solution}

\begin{figure}[t]
    \centering
    \includegraphics[trim={0 0 0 0},clip,width=\textwidth]{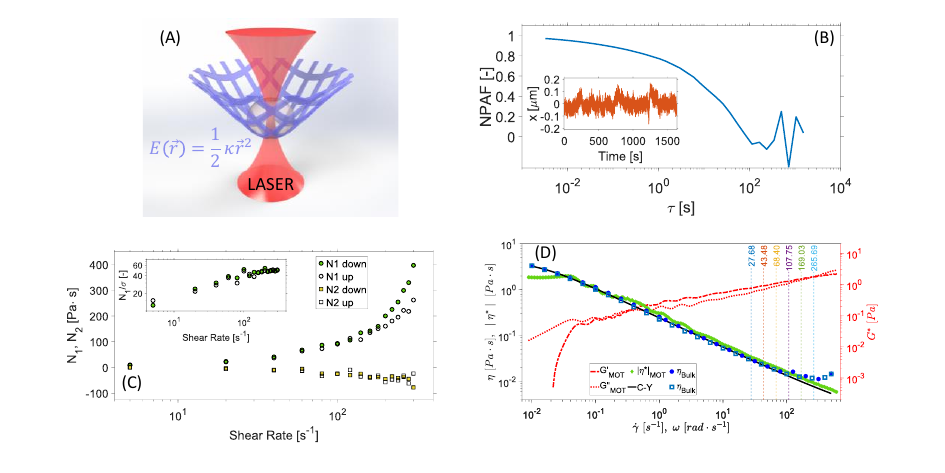}
    \caption{(A) A schematic representation of an optically trapped bead within a harmonic potential, $E(\vec{r})$, where $\kappa$ is the trap stiffness and $\vec{r}$ is the bead position from the trap centre.
    (B) The normalised position autocorrelation function (NPAF) \textit{versus} lag-time ($\tau$) calculated using the x coordinate of the trajectory of an optically trapped bead suspended in a $0.07\%$wt PAM solution shown in the inset. 
    (C) The first and the second normal stress differences ($N_1$ and $N_2$, respectively) obtained using a cone and plate and parallel plate configurations respectively \textit{versus} the shear rate. The inset displays the ratio between $N_1$ and the shear stress ($\sigma$) \textit{versus} the shear rate. These measurements have been corrected for inertia which did not exceed $10$\% of the signal.
    (D) The viscoelastic moduli (red lines, $G'(\omega),G''(\omega)$) and the complex viscosity (green diamond, $|\eta^*(\omega)|_{MOT}$) as obtained from the analysis of the NPAF shown in (B). In the same figure the shear viscosity (blue symbols, $\eta$) and its fit by means of Carreau-Yasuda model (black line, C-Y) are reported; $\eta$ and $|\eta^*(\omega)|_{MOT}$ are plotted \textit{versus} shear rate ($\dot{\gamma}$) and angular frequency ($\omega$), respectively; in compliance to the Cox-Merz rule. The vertical dotted lines represent the shear rates explored in the shear cell experiments (see table \ref{tab:ShearRate}).}
    \label{fig:MicroPAM}
\end{figure}

The rheological properties of a PAM solution at a concentration of $0.07\%$wt were measured by means of both microrheology measurements performed with optical tweezers and conventional rotational shear rheology, as summarised in Fig.\ref{fig:MicroPAM}-(D). In the case of microrheology, the trajectory of an optically trapped particle was captured at circa $1$kHz for an extended measurement duration of circa $27$ minutes, in compliance to the arguments described in detail by Smith \textit{et al.}\cite{smith2023machine} to achieve statistically valid outcomes. The NPAF of the particle trajectory was analysed by using i-Rheo MOT \cite{Tassieri2012}, an algorithm based on the analytical method developed by Evans \textit{et al}.\cite{evans2009direct} for evaluating the Fourier transform of any generic function, sampled over a finite time window, without the need for Laplace transforms or fitting functions. For more detail about the principles underpinning i-Rheo MOT, the reader is advised to read these references \cite{tassieri2016rheo,Smith2021,Tassieri2019,Tassieri2018}.

The first and the second normal stress differences ($N_1$ and $N_2$, respectively) for the $0.07\%$wt PAM solution are shown in Fig.\ref{fig:MicroPAM}-(C). They are in good agreement with the theoretical predictions of uniform shear flow of inelastic hard spheres in dilute regime, for which $N_1$ is expected to be positive and monotonically increasing for higher shear rates; while, $N_2$ is expected to be negative and monotonically decreasing for higher shear rates. Additionally, in the inset of Fig.\ref{fig:MicroPAM}-(C), we report the ratio between $N_1$ and the applied shear stress ($\sigma$), which shows a plateauing behaviour at shear rates close to the onset of the shear induced elastic instability.

In Fig.\ref{fig:MicroPAM}-(D), we report the viscoelastic moduli ($G'(\omega),~G''(\omega)$) and the complex viscosity ($|\eta^*(\omega)|_{MOT}$) as obtained by means of MOT measurements together with the rotational shear viscosity ($\eta_{Bulk}$) obtained by using a stress-controlled rheometer.
Strikingly, despite the substantial difference in the nature of the two rheological techniques employed here, the outcomes of these experimental approaches are in very good agreement over a range of shear rates/frequencies spanning $\sim 5$ decades.
Moreover, it is important to highlight that, while at high shear rates (i.e., $\dot{\gamma}>107.75$s$^{-1}$), bulk rheology measurements reveal the onset of the elastic instability as inferred by the blue symbols departing from the Carreau-Yasuda fit (black line) of the flow curve in a shear thickening behaviour; this phenomenon is not revealed by the complex viscosity curve because of the quiescent nature of MOT measurements.
Nonetheless, the viscoelastic moduli reveal the existence of two characteristic crossover frequencies occurring at (I) circa $0.3$rad/s and (II) circa $200$rad/s. The inverse of these frequencies provide a measure of two of the material's characteristic relaxation times, i.e.: the reptation time $\tau_{rep}$ and the entanglement time $\tau_{e}$, respectively.
These measurable parameters can be used to educe the material's Rouse time ($\tau_R$) \cite{colby2003polymer}:
\begin{equation}
    \tau_R = \tau_e\Bigg( \frac{\tau_{rep}}{6\tau_{e}}\Bigg)^{2/3}\cong 0.12~sec,
    \label{eq:Rouse}
\end{equation}
which, as we shall demonstrate hereafter, correlates very closely with the instability time~\cite{howe2015flow} derived from bulk rheology measurements.

The vertical lines in Fig.\ref{fig:MicroPAM}-(D) are the experimental shear rates investigated by using the cone and plate shear cell (see table \ref{tab:ShearRate}). These shear rates span the range of the shear induced elastic instability and cover its onset, with the aim to provide new insights into FISA under this condition.
At this point it is important to remind that, the onset of elastic instability has been described by means of a dimensionless parameter $M$, first introduced by McKinley \textit{et al.}\cite{mckinley1996rheological,pakdel1996elastic}, which is related to both the Weissenberg and the Deborah numbers of the system under investigation:
\begin{equation}
    M = \sqrt{Wi\cdot De}.
\end{equation}
This equation has been made explicit in the case of a cone and plate geometry\cite{howe2015flow}, and here further modified for a counter rotating configuration:
\begin{equation}
    M_{Norm} = \frac{M}{M_{crit}} = \frac{\lambda_{PM}(\omega_c+\omega_p)}{\sqrt{\theta}}.\frac{1}{M_{crit}},
\end{equation}
where $\lambda_{PM}$ is a characteristic relaxation time of the viscoelastic fluid, for which we have adopted the suffix $PM$ for Pakdel-Mckinley \cite{pakdel1996elastic} to distinguish this time as already done by Howe \textit{et al.}\cite{howe2015flow}, and $M_{crit}$ is the critical value of $M$ when the flow becomes unstable. Interestingly, a linear-stability analysis~\cite{howe2015flow,pakdel1996elastic,olagunju1995instabilities} of a cone and plate geometry has shown that flow instability initiates at $M \ge 21.17$. Thus, by using this inequality and the onset of shear induced instability in Fig.\ref{fig:MicroPAM}-(D) at 126.5s$^{-1}$, it is possible to determine a characteristic relaxation time $\lambda_{PM}$ of $0.14$s, which can be associated to the material's Rouse time as described by Howe \textit{et al.}~\cite{howe2015flow} and here further corroborated by the value obtained from microrheology measurements with optical tweezers via Equation~\ref{eq:Rouse}.

\subsection*{Accumulation of Particles}

\begin{figure}[ht]
    \centering
    \includegraphics[clip,width=0.5\textwidth]{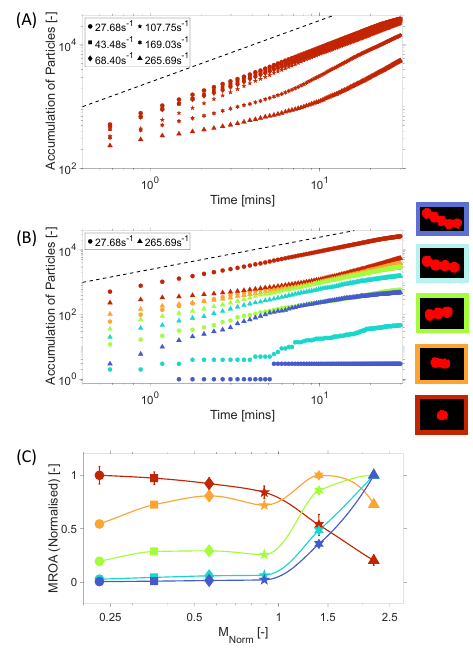}
    \caption{(A) Accumulation of particles for a chain length of 1 \textit{versus} time for shear rates ranging from $27.68$s$^{-1}$ to $265.69$s$^{-1}$. Each curve has been down-sampled for easier identification and the dashed black line indicates a linear growth. (B) Accumulation of particles for each chain length \textit{versus} time for shear rates of $27.68$s$^{-1}$ and $265.69$s$^{-1}$. As in (A) each curve has been down-sampled for easier identification and the dashed black line indicates a linear growth. (C) Mean rate of accumulation (MROA), normalised by the maximum MROA for each particle chain length \textit{versus} $M_{Norm}$. The colours associated with each curve represent a particle chain length as for by the outline of the single, dimer, trimer, tetra and penta images on the right side. The outputs shown in this figure were obtained in triplicates.}
    \label{fig:Accumulation}
\end{figure}

FISA phenomenon can be analysed by calculating the accumulation of chains, of a given length, over each successive image frame. The analysis can be thought of as a series of `bins' into which different chain lengths are added. The number of chains in each bin can easily be summed over time to produce information on the long time effect of shear rate on the chaining occurring within the sample. As mentioned earlier, the image analysis software separated the particles/chains identified in each image frame into five different `bins', i.e.: singles, dimers, trimers, tetras and pentas. The curves shown in Fig.\ref{fig:Accumulation}-(A) are the total number of single particles accumulated throughout the experiment for each imposed shear rate. Whereas, in Fig.\ref{fig:Accumulation}-(B) we report the accumulation of each chain length for the two extreme shear rates explored in this work.

From Fig.\ref{fig:Accumulation}-(A-B) it is clear and expected that the accumulation of particles increases with time. However, it is interesting to notice that the curves settle into clear ``bands'' depending on the relative value of $M_{Norm}$. Indeed, from Fig.\ref{fig:Accumulation}-(A) it is apparent that the accumulation curves overlay on each other at relatively low shear rates (i.e., for $\dot{\gamma} \leq 107.75$s$^{-1}$ or equivalently for $M_{Norm}<1$), but they branch off at higher ones (i.e., for $\dot{\gamma} \geq 169.03$s$^{-1}$ or equivalently for $M_{Norm}>1$).
The differences between the curves representing the accumulation of particles is most apparent for the dark blue symbols in Fig.\ref{fig:Accumulation}-(B), which are representative of the pentas at the smallest and the largest explored shear rates, i.e. $27.68$s$^{-1}$ and $265.69$s$^{-1}$, respectively. Here, the curve at the largest shear rate (denoted by triangular symbols) has a significantly greater accumulation of particles than the one at lowest shear rate (circle symbols); thus, implying a shear induced enhancement of the generation of pentas.
Interestingly, this phenomenon is better explicated by the mean rate of accumulation (MROA) of the curves as reported in Fig.\ref{fig:Accumulation}-(C); which has been determined by normalising the mean value of the time derivative of the accumulation curves by its maximum value. In Fig.\ref{fig:Accumulation}-(C) the MROA is reported against $M_{Norm}$, and it can be observed that for single particles the MROA decreases with increasing $M_{Norm}$; whereas, for dimers it stays relatively constant across the same range of explored $M_{Norm}$. In contrast, the MROA of the longer chain particles has a relatively low and constant value for $M_{Norm}<1$, with a sharp increase for $M_{Norm}>1$.
At this point it is important to remind that, in these experiments the total number of individual polystyrene beads is constant, and therefore as chains of different length start to form, the number of single particles decreases.
It follows that, at low $M_{Norm}$, particles are not able to form long chains and the MROA value of single particles is high. Whereas, as the shear rate increases to a point where the $M_{Norm}$ is greater than $1$, the elastic instability of the PAM solution enhances the chaining, which results in a significant drop of the MROA value for the single particles, while the MROA of the trimers, tetras and pentas increases rapidly.

\subsection*{Relative Population of Chains}
\begin{figure}[ht]
    \centering
    \includegraphics[trim={0 0 0 0},clip,width=\textwidth]{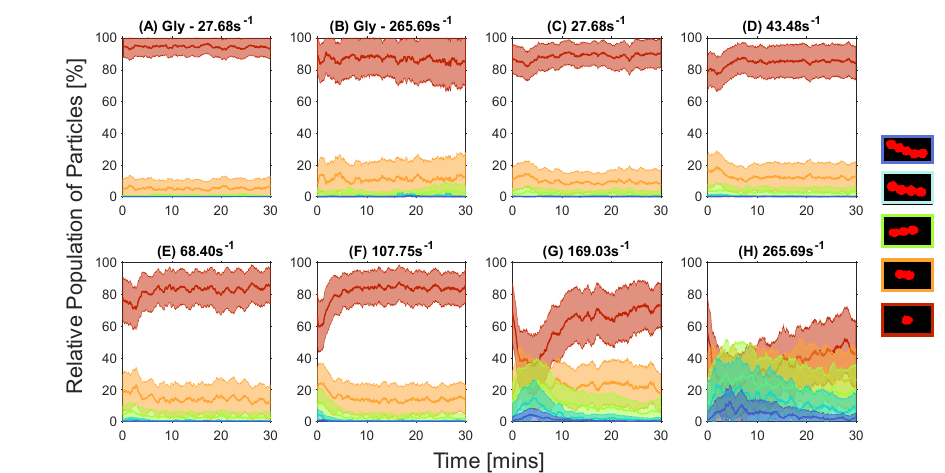}
    \caption{Relative population of particle chains (singles, dimers, trimers, tetras and pentas) \textit{versus} time. (A and B) Particle suspension in glycerol/water mixture at a shear rates of  $27.68$s$^{-1}$ and $265.69$s$^{-1}$, respectively. (C-H) PAM solutions at shear rates spanning from $27.68$s$^{-1}$ to $265.69$s$^{-1}$, with related values reported in table \ref{tab:ShearRate}, respectively. The shaded areas are the standard deviation associated with the experimental triplicates and each curve has been smoothed by using a moving average window of $30$s.}
    \label{fig:Percentage}
\end{figure}

A complementary method to analyse the progression of FISA within the sample at different shear rates, is by identifying the relative population of each chain length in the image frame. To achieve this, the percentage of particles was calculated by taking the ratio between each chain length and the total number of particles identified in each frame, as shown in Fig.\ref{fig:Percentage}. 

In particular, in Fig.\ref{fig:Percentage}-(A-B) are reported the relative population of particle chains for suspensions made with glycerol/water mixture and measured at the two extremes of the range of explored shear rates (i.e., $27.68$s$^{-1}$ and $265.69$s$^{-1}$). From these diagrams it is apparent that no significant changes occur and that the mean percentage of single particles identified in the image frames stays well above $80\%$; whereas, for all the other particle chain sizes, it remains well below $20\%$. Thus confirming that chaining does not occur in Newtonian fluids.
Interestingly, this is not the case of particle suspensions in the viscoelastic solution employed in this work. Indeed, as shown in Fig.\ref{fig:Percentage}-(C-F), there is an initial fall of the mean percentage of single particles whose magnitude increases with the applied shear rate. This is followed by a corresponding rise in the mean percentage of longer chains, all within the first $5$ minutes from the start of the experiments. Then the percentage of single particles begins to climb back to an almost steady value for the duration of the experiment, which is higher than $80\%$ for $M_{Norm}<1$ and lower than $80\%$ for $M_{Norm}> 1$.
A similar, but opposite behaviour is seen for the longer particle chains, where an initial increase of their population is observed over the same time scale (i.e., $5$min), followed by a decrease towards a steady value.
One could argue that, a possible explanation of such dynamic process could be related to the migration of single particles and chains from/into the focal plane; however, the complementarity of these processes between the single particles and the chains (i.e., the decrease/increase in single particles is complemented by the increase/decrease in the percentage of dimers, trimers, tetras and pentas) implies (i) that the total number of particles is constant during the measurement and (ii) that longer chain particles are breaking down back to single particles.  

\subsection*{Alignment Factor}
An additional method to analyse the progression of FISA is by means of the alignment factor ($A_f$), as described by Pasquino \textit{et al}. \cite{pasquino2013migration}, which is defined as:
\begin{equation}
    A_f = \frac{\sum_{L=1}^{L_{max}}N_LL^2}{\sum_{L=1}^{L_{max}}N_LL},
\end{equation}
where $L$ is the chain length and $N_L$ is the number of chains of a given length $L$ in an image frame.
As suggested by its name, $A_f$ is a measure of bead alignment in a given image frame and will always have a value $\geq 1$, where $A_f = 1$ is achievable \textit{if and only if} there were solely single beads in the image frame. As stated by Pasquino \textit{et al}. \cite{pasquino2013migration}, $A_f$ bares a resemblance to the weight average molecular weight of polymer chains and as such chains of longer length have greater impact on $A_f$. 

\begin{figure}[ht]
    \centering
    \includegraphics[trim={0 0 0 0},clip,width=\textwidth]{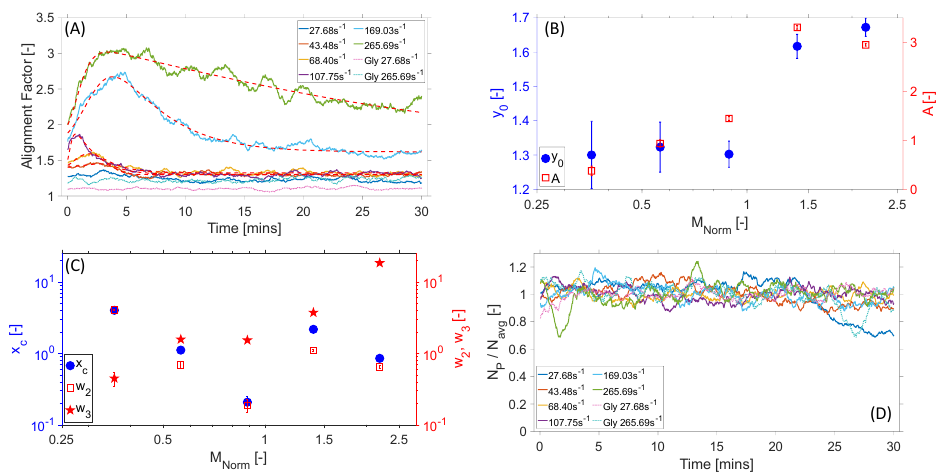}
    \caption{(A) Alignment factor curves (averaged over triplicates) for the explored shear rates, including the two set of measurements performed on the $97\%$ glycerol/water mixture (Gly) as control (dotted lines). The red dashed lines represent Eq.\ref{eqn:asym2sig} as fitting function. Notice that, each curve has been smoothed by using a moving average window of $60$s width. (B) The resulting fitting parameters $y_0$ and $A$ of Eq.\ref{eqn:asym2sig} \textit{versus} $M_{Norm}$. (C) The resulting fitting parameters $x_c$ (left axis), $w_2$ and $w_3$ (right axis) of Eq.\ref{eqn:asym2sig} \textit{versus} $M_{Norm}$. The standard deviation of the fitting function is depicted as error bars in (B) and (C). (D) The normalised number of particles \textit{versus} time, used to analyse the migration of particles away from the focal plane. Here, each curve has been smoothed by using a moving average window of $30$s width.}
    \label{fig:AlignmentFactor}
\end{figure}

In Fig.\ref{fig:AlignmentFactor}-(A) we report the alignment factor for the same set of experiments analysed earlier in Fig.\ref{fig:Accumulation} and \ref{fig:Percentage}.
From Fig.\ref{fig:AlignmentFactor}-(A) it can be seen that at relatively short times, all curves show an increase of $A_f$ up to a maximum value, whose amplitude increases proportionally to the shear rate, while its abscissa is inversely proportional to $\dot{\gamma}$ for $M_{Norm}<1$ and remains almost constant for $M_{Norm}> 1$. After reaching a maximum, all the curves tend to exponentially decrease towards a steady-state value at long times.
This is the first time in literature that such behaviour of $A_f$ is reported, as it was expected that $A_f$ would have increased monotonically until a plateau value at long times, as described by Pasquino \textit{et al}. \cite{pasquino2013migration,pasquino2014}. However, it must be said that, their focal plane was positioned at the wall of the shear cell and therefore particle migration\cite{pasquino2010effect,pasquino2013migration} towards the plate continued to supply the area with new beads. In this work, the focal plane was placed at the centre of the fluid chamber and therefore we again posit that the decrease of $A_f$ at long-times may be due to either (i) particle migration away from the centre of the channel (this being a possible explanation that is not supported by the experimental evidence reported in this work), or (ii) longer chain particles breaking down to single particles (which is the thesis we support).

To better understand the temporal behaviour of $A_f$, we performed a best fit of the curves by means of the following asymmetric double sigmoid function\cite{chen2017thermogravimetric} (also known as piecewise logistic function):
\begin{equation}
    y = y_0 + A\frac{1}{1+e^{-\frac{x-x_c+w_1/2}{w_2}}}\left(1-\frac{1}{1+e^{-\frac{x-x_c-w_1/2}{w_3}}}\right),
    \label{eqn:asym2sig}
\end{equation}
where $y_0$ is the steady state value, $A$ is the amplitude, $x_c$ is the peak center, $w_1$ is the curve width, $w_2$ and $w_3$ are shape parameters.
For the fitting curves shown in Fig.\ref{fig:AlignmentFactor}-(A), it was found that $w_1=0$ for all fits; whereas, the remaining parameters were different from zero and they are plotted against $M_{Norm}$ in Fig.\ref{fig:AlignmentFactor}-(B) and (C). 
From these figures, it can be seen that $y_0$ is almost constant for $M_{Norm}<1$, with a significant increase for $M_{Norm}>1$. Whereas, the amplitude $A$ shows a progressive increase with $M_{Norm}$; although, it could be argued that there is an initial shallow increase for $M_{Norm}<1$ and then a relatively sharp increase for $M_{Norm}>1$.
From Fig.\ref{fig:AlignmentFactor}-(C) one will notice that $x_c$ (the peak centre) behaves rather erratically when plotted against $M_{Norm}$. Indeed, it initially decreases for $M_{Norm}<1$ (which corresponds to a reduction in time for the peak to occur in Fig.\ref{fig:AlignmentFactor}-(A)); however, as $M_{Norm}$ exceeds $1$, there is first a sharp increase in $x_c$ and then it starts decreasing again.
Interestingly, a similar behaviour is shown by $w_2$.
Finally, $w_3$ is the `shape parameter' for the curve after its peak, and an increase in its value would suggest an increase of the characteristic ``\textit{relaxation time}'' of $A_f$. Thus suggesting that the degradation of the particle chains is controlled by a different physical process than the one governing the generation of particle chains.

Overall, the above analysis of the fitting parameters indicates a clear change in behaviour of the particle chaining phenomena as $M_{Norm}$ exceeds $1$; with a sharp enhancement of the chaining that correlates well with the onset of the elastic instability of the fluid.

\subsection*{Flow induced particle chains break down}

Both the analysis of the relative chain length population and the $A_f$ have raised the same question: is the decrease in FISA at long-times due to the migration of longer chain particles away from the focal plane or is it due to the break down of these same particle chains back to single particles? 

To address this question, we have followed the approach introduced by Mirsepassi \textit{et al}. \cite{mirsepassi2012particle}, whereby one could reveal the existence of migration during flow by normalising the change in the number of particles in each frame (independently on whether they are single or belonging to chains of different length) to the average number of particles in each image determined over the entire measurement duration. In Fig.\ref{fig:AlignmentFactor}-(D) we report the results of such analysis with a moving average window of $30$s width. Despite the existence of significant fluctuations, all curves fluctuates around a constant value equal to $1$, which suggests that the average number of particles in the bulk of the PAM solution does not change during the ($30$mins) measurements; 
hence, it is unlikely that the migration of particles out of the focal plane can fully justify the decrease in $A_f$, although it cannot be fully discarded. Additionally, the sedimentation time of the particles was considered, due to the decrease in the $27.68$s$^{-1}$ PAM curve after $\sim20$mins, but no agglomerates were observed at the bottom of the shear cell after the experiments, suggesting its influence is insignificant and therefore the decrease seen in this curve is caused by a yet unknown phenomenon. 
These results further corroborate our hypothesis that the reduction in FISA at long-times is likely to be due to the longer particle chains breaking down back into shorter chain lengths.

This poses a further question: what would cause the particles chains to break down? A possible cause would involve the flow induced disentanglement of the polymer chains, as described by Vasquez \textit{et al.} \cite{vazquez2001shear}. This phenomenon would occur on time scales much longer than the characteristic relaxation time ($\lambda$) and would produce a drop in the viscoelastic properties of PAM solutions during shear; thus dimming the elastic character of the fluid and therefore potentially reversing the particle chaining (as in Fig.\ref{fig:AlignmentFactor}-(A)) by shifting the $M_{Norm}$ towards values lower than one.
Interestingly, this thesis is supported by the experimental evidences reported in a recent study by Tran and Clarke \cite{tran2023relaxation}, where they performed parallel-superposition shear measurements to reveal (i.e., see Figure $2$ of their study) a significant reduction of the longest characteristic relaxation time of high molecular weight PAM solutions as function of the imposed shear rate. These results unveil a shift of the low-frequency crossover of the viscoelastic moduli towards higher frequencies, which would imply a reduction of the fluid's longest characteristic relaxation time. However, the results reported by Tran and Clarke \cite{tran2023relaxation} were recorded at steady state and therefore they do not inform on the dynamics of the above phenomenon, which one would expect to be potentially correlated to the characteristic time-scales shown in Fig.\ref{fig:AlignmentFactor}.

\begin{figure}[ht]
    \centering
    \includegraphics[trim={0 0 0 0},clip,width=8cm]{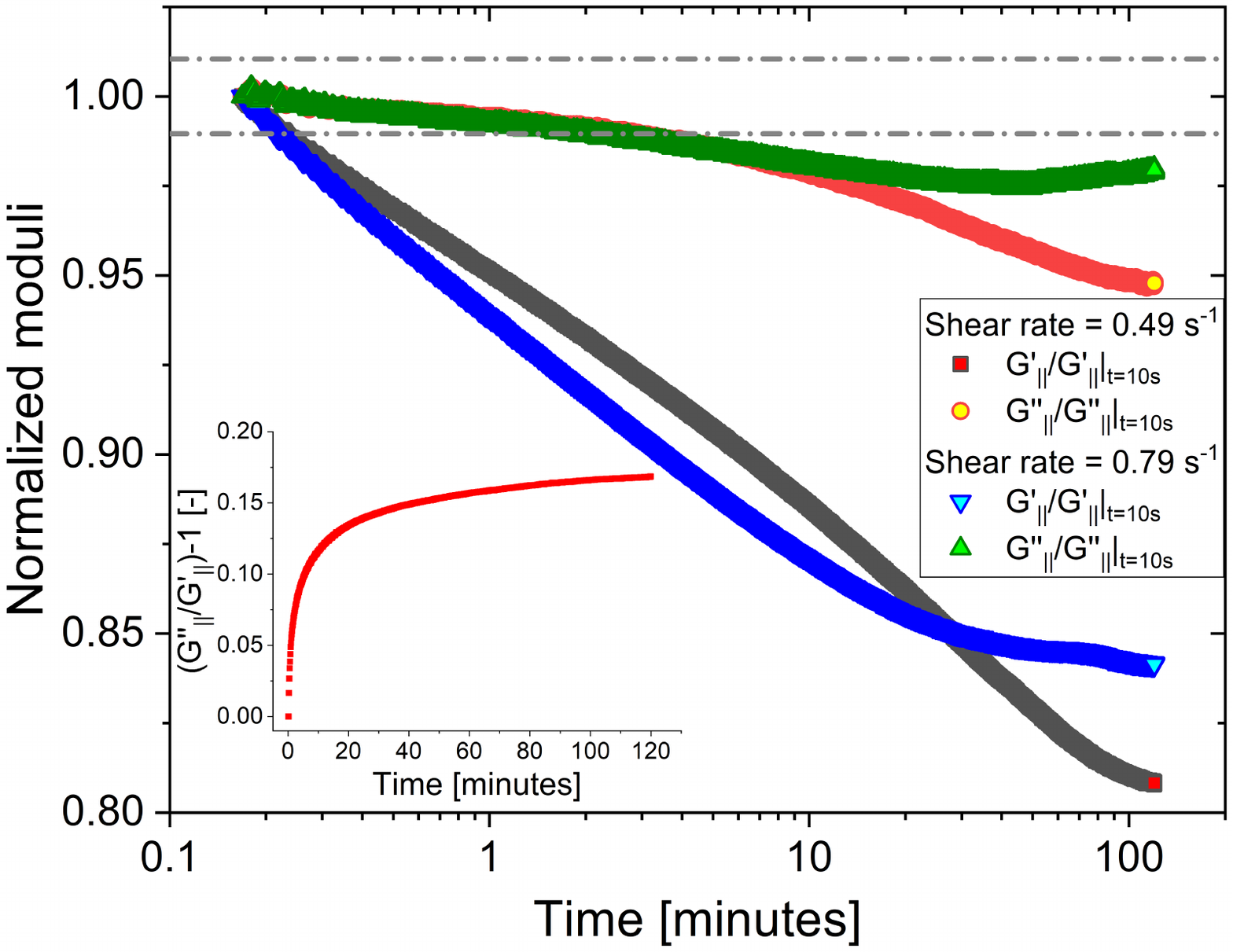}
    \caption{(Symbols) The normalized viscoelastic moduli \textit{versus} time of a PAM solution at concentration of $0.07\%$wt as measured by means of two parallel superposition shear measurements performed at shear rates of $0.49$s$^{-1}$ and $0.79$s$^{-1}$, respectively. The moduli were measured at a constant frequency of $0.4$Hz and normalized by their value at time $10$s. The two dashed lines indicate the values of $1\pm$ standard deviation of the viscoelastic moduli (measured at a constant frequency of $0.4$Hz) \textit{versus} time of the same solution as above, but at zero shear (i.e., as obtained from a time-sweep of a small amplitude oscillatory measurement); in this case the moduli were normalized by their mean values, respectively. The inset shows the averaged ratio of the viscous modulus to that of the elastic modulus minus one for the two set of measurements shown in the main.}
    \label{fig:SRS}
\end{figure}

In Fig.\ref{fig:SRS} we report, for the first time in literature, an experimental evidence revealing the time scale of the phenomenon described above. In particular, Fig.\ref{fig:SRS} shows the normalized viscoelastic moduli \textit{versus} time of a PAM solution at concentration of $0.07\%$wt as obtained by two parallel superposition shear measurements performed at shear rates of $0.49$s$^{-1}$ and $0.79$s$^{-1}$, respectively. The measurements were performed at a constant frequency of $0.4$Hz and the moduli normalized by their value at time equal to $10$s (i.e., $G'_{||}/G'_{||}|_{t=10}$ and $G''_{||}/G''_{||}|_{t=10}$). The two lines indicate the values of $1\pm$ standard deviation of the viscoelastic moduli \textit{versus} time of the same solution as above, but measured at zero shear (i.e., as obtained from a time-sweep of a small amplitude oscillatory measurement); in this case the moduli were normalized by their mean values, respectively.

From Fig.\ref{fig:SRS} it is possible to argue that: (i) there is a significant reduction over the time of the elastic component of the polymer solution (i.e., $G'_{||}\propto t^{-1/15}$), when the fluid is subjected to a continuous shear rate (thus supporting the thesis of polymers disentanglement); (ii) the viscous modulus is significantly less affected than the elastic component (because the polymer concentration and its molecular conformation are almost constant during the measurement); (iii) most of the significant changes of the viscoelastic properties of the solution occur within the first $\sim 20$ minutes of the measurement (as revealed by the data shown in the inset), which is in very good agreement with the time-scales of the alignment factor shown in Fig.\ref{fig:AlignmentFactor}, especially for those measurements having the same shear rates.

\section*{Conclusions} 
In this work we have studied the flow-induced self-assembly (FISA) of particles suspended into both Newtonian and viscoelastic fluids. This has been achieved by means of a bespoke shear cell that has allowed us to monitor the dynamics of the particle chain `generation' and for the first time in literature `degradation'. 
In particular, the adoption of a viscoelastic fluid characterised by a clear transition between a shear thinning behaviour (i.e., with a decreasing viscosity) at relatively low frequencies (or equivalently at low shear rates) and a shear-thickening character (onset of elastic instability) at relatively high frequencies (or shear rates) and the fine tuning of the applied shear rates, have allowed us to investigate the particle chaining phenomenon across a $M_{Norm}$ of $1$. In particular, this study has led to the following key findings:
(i) we have corroborated that particles suspended into Newtonian fluids do not show FISA enhancement at different shear rates;
(ii) FISA within viscoelastic fluids is significantly greater than in Newtonian fluids, when compared over the same range of shear rates;
(iii) for $M_{Norm}$ higher than $1$, the onset of the elastic turbulence of the viscoelastic fluid correlates strongly with the enhancement of FISA;
(iv) particle chains break apart when a constant shear is applied for sufficiently long-time (i.e. much longer than the fluids' longest relaxation time).
(v) We provide for the first time in literature experimental evidence of a significant reduction of the fluid's elastic character when a continuous flow is applied for sufficiently long time.
Points (iv) and (v) have been corroborated (via private communication) by means of computational fluid dynamics simulations kindly performed by Pier Luca Maffettone and Gaetano D’Avino (data not reported here), whose outcomes will be published in a separate publication.

Finally, we envisage that, in future works the adoption of imaging systems with a wider field of view capable of capturing images of the whole shear cell at higher resolution and acquisition rate, would allow a deeper understanding of the dynamics of both FISA and chain breakdown phenomena. Moreover, further investigation of the rheological properties of PAM solutions by means of parallel superposition shear measurements, may elucidate the newly discovered phenomenon of the reduction of the fluid's viscoelastic properties when subjected to a continuous flow for sufficiently long time.

\section*{Acknowledgements}
The authors thank Rossana Pasquino and Francesco Del Giudice for the informative discussions. We thank Pier Luca Maffettone and Gaetano D’Avino for corroborating our experimental results by means of computational fluid dynamics simulations. This work was supported by the EPSRC CDT in Intelligent Sensing and Measurement, Grant Number EP/L$016753$/$1$.

\section*{Author contributions statement}

M.T. conceived the experiment(s), M.G.S. conducted the experiments with the shear cell, M.G.S. and M.T. analysed the data, M.G.S. wrote the manuscript, G.M.G. \& M.T. conducted the microrheology measurements, T.F. \& A.L. supervised the shear cell measurements, A.C. \& A.R. conducted the bulk rheology measurements. All authors contributed to the drafting of the manuscript.

\bibliography{References}

\end{document}